\documentclass{aa}
\usepackage{times,graphicx}
\usepackage{curves,epic,eepic}

\begin{document}

   \thesaurus{06           
              (13.07.1;)  
              (11.19.1;)} 

            \title{An optical study of the GRB 970111 field beginning 19
              hours after the Gamma-Ray Burst~\thanks{Based on observations
                collected at the German-Spanish Astronomical Center, Calar
                Alto, operated by the Max-Planck-Institut f\"ur Astronomie,
                Heidelberg, jointly with the Spanish National Commission
                for Astronomy.}$^{,}$\thanks{Based on observations carried
                out at the Danish 1.54-m Telescope on the European Southern
                Observatory, La Silla, Chile}$^{,}$\thanks{Based on
                observations at the Osservatorio Astronomico di Loiano,
                Italy.}}

   \author{J. Gorosabel
          \inst{1}
   \and A. J. Castro-Tirado
          \inst{1,2}
   \and C. Wolf
          \inst{3}
   \and J. Heidt
          \inst{4}
   \and T. Seitz
          \inst{4,5}
   \and E. Thommes
          \inst{6}
   \and C. Bartolini
          \inst{7}
   \and A. Guarnieri
          \inst{7}
   \and N. Masetti 
          \inst{7,8}
   \and A. Piccioni
          \inst{7}
   \and S. Larsen
          \inst{9}
   \and E. Costa
          \inst{10}
   \and M. Feroci
          \inst{10}
   \and F. Frontera
          \inst{11}
   \and E. Palazzi
          \inst{8}
   \and N. Lund
          \inst{12}
           }

   \offprints{J. Gorosabel (jgu@laeff.esa.es)}

   \institute{ Laboratorio de Astrof\'{\i}sica Espacial y F\'{\i}sica
     Fundamental (LAEFF-INTA), P.O. Box 50727, E-28080 Madrid, Spain.
\and Instituto de Astrof\'{\i}sica de Andaluc\'{\i}a (IAA-CSIC), P.O. Box
     03004, E-18080 Granada, Spain.
\and Max-Planck-Institut f\"ur Astronomie, Heidelberg, Germany. 
\and Landessternwarte Heidelberg, K\"onigstuhl, 69117 Heidelberg, Germany.
\and Lehrstuhl f\"ur Ergonomie, Technische Universit\"at M\"unchen, 
Boltzmannstr. 15, 85747 Garching, Germany.
\and Royal Observatory, Blackford Hill, Edinburgh Hill, Edinburgh EH9 3HJ,
     United Kingdom.
\and Dipartimento di Astronomia, Universit\`a di Bologna, Via
              Zamboni  33, I-40126 Bologna, Italy. 
\and Istituto Tecnologie e Studio Radiazioni Extraterrestri, CNR, Bologna,
     Italy. 
\and Niels Bohr Institute for Astronomy, Physics and Geophysics, Copenhagen
     University Astronomical Observatory, Julian Maries Vej 30, 2100 
     Copenhagen, Denmark. 
\and Istituto di Astrofisica Spaziale, CNR, Roma, Italy. 
\and Dipartamento di Fisica, Universit\`a di Ferrara, Ferrara, Italy. 
\and Danish Space Research Institute, Copenhagen, Denmark.}
         
   \date{Received date; accepted date}

   \titlerunning{Study of the GRB 970111 field.}

  \authorrunning{Gorosabel et al.}

   \maketitle

   \begin{abstract} We present the results of the monitoring of the GRB
     970111 field that started 19 hours after the event. This observation
     represents the fastest ground-based follow-up performed for GRB 970111
     in all wavelengths. As soon as the detection of the possible GRB
     970111 X-ray afterglow was reported by Feroci et al.  (1998) we
     reanalyzed the optical data collected for the GRB 970111 field.
     Although we detect small magnitude variability in some objects, no
     convincing optical counterpart is found inside the WFC error box. Any
     change in brightness 19 hours after the GRB is less than 0.2 mag for
     objects with B $<$ 21 and R $<$ 20.8.

     The bluest object found in the field is coincident with 1SAX\-
     J1528.8\-+1937.  Spectroscopic observations revealed that this object
     is a Seyfert-1 galaxy with redshift $z=0.657$, which we propose as the
     optical counterpart of the X-ray source.

     Further observations allowed to perform multicolour photometry for
     objects in the GRB 970111 error box.  The colour-colour diagrams do
     not show any object with unusual colours. We applied a photometric
     classification method to the objects inside the GRB error box, that
     can distinguish stars from galaxies and estimate redshifts. We were
     able to estimate photometric redshifts in the range $0.2 < z < 1.4$
     for several galaxies in this field and we did not find any conspicuous
     unusual object.

     We note that GRB 970111 and GRB 980329 could belong to the same class
     of GRBs, which may be related to nearby sources ($z \sim 1$) in which
     high intrinsic absorption leads to faint optical afterglows.


   \end{abstract}

%

\section{INTRODUCTION}
Gamma-Ray Bursts (GRBs) are powerful brief transient phenomena of
high-energy radiation that appear randomly in the sky. They were first
detected in 1969 by the {\it Vela} satellites (Klebesadel et al. 1973) and
have remained for many years one of the most elusive mysteries in
astrophysics.

Before the launch of the {\it BeppoSAX} and {\it RossiXTE} sate\-llites in
1996, they had not been detected in any other wavelength region, and their
distance scale remained unknown. The discovery of X-ray afterglows by both
satellites revolutionized the field, because they are able to provide very
accurate GRB error boxes within a few hours (circular error boxes with
radius up to $50^{\prime \prime}$), which enables very rapid follow up
observations at longer wavelengths. Eight GRBs detected by {\it BeppoSAX}
have been detected at optical and infrared wavelengths; GRB 970228
(Guarnieri et al. 1997a, van Paradijs et al. 1997), GRB 970508 (Bond 1997,
Djorgovski et al.  1997, Castro-Tirado et al.  1998a), GRB 971214 (Halpern
et al.  1997, Gorosabel et al. 1998, Rama\-prakash et al. 1998), GRB 980326
(Groot et al.  1998), GRB 980329 (Klose 1998, Palazzi et al.  1998a, 1998b,
Taylor et al. 1998), GRB 980519 (Jaunsen et al. 1998, Maury et al.  1998),
GRB 980613 (Hjorth et al. 1998, Djorgovski et al. 1998a, 1998b), and GRB
980703 (Frail et al.  1998, Zapatero-Osorio et al. 1998).

Spectral information taken for GRB 970508 showed that the gamma event
occurred at a redshift $z\geq$0.83, therefore supporting the cosmological
origin of GRBs (Metzger et al. 1997). Recently, Kulkarni et al. (1998)
measured the redshift of the host-galaxy of GRB 971214 ($z$=3.42), placing
it at a distance of $\sim$12 billion light years (assuming the universe to
be about 14 billion years old). This large distance implies an energy
release at least of $\sim 3 \times 10^{53}$ erg, which can not be explained
by the neutron star merger model, at least in its simplest form (Narayan et
al. 1992).

 \begin{figure}[t]
\fbox{\rotatebox[origin=c]{-90}{\resizebox{8.2cm}{8.6cm}{\includegraphics{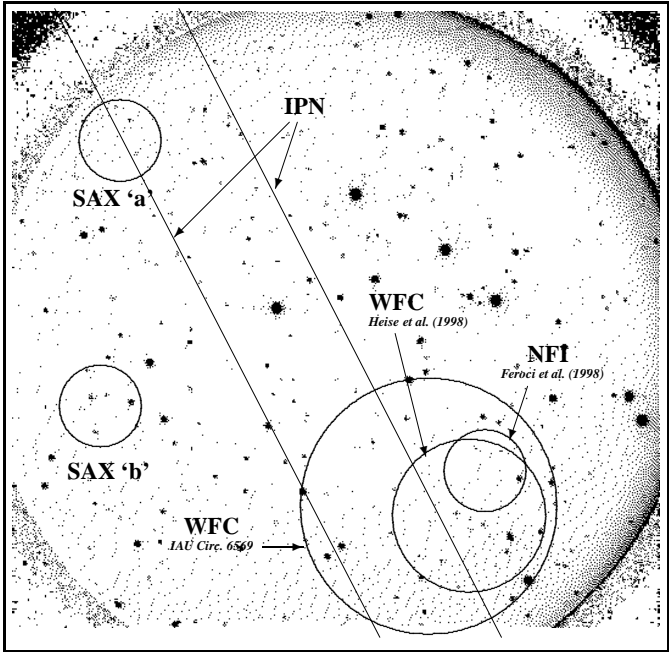}}}
\put(-50,106) {\bfseries NFI}
\put(-60,100){\itshape \bfseries \tiny Feroci et al. (1998)}
\put(-50,97){\vector(-1,-3){6.8}}
\put(-110,126) {\bfseries WFC}
\put(-110,120){\itshape \bfseries \tiny Heise et al. (1998)}
\put(-100,117){\vector(1,-4){11.7}}
\put(-180,41) {\bfseries WFC}
\put(-185,35){\itshape \bfseries \tiny IAU Circ. 6569}
\put(-150,36){\vector(1,0){15.3}}
\put(-142,200) {\bfseries  IPN}
\put(-140,197){\vector(-1,-1){43.6}}
\put(-137,197){\vector(-1,-3){8.4}}
\put(-222,165){\bfseries SAX `a'}
\put(-224,62) {\bfseries SAX `b'}
\put(-87.3,51.6){\bigcircle{97}}
\put(-72,48){\bigcircle{58}}
\put(-66,65){\bigcircle{31}}
\put(-211.5,89.5){\bigcircle{31}}
\put(-204,190){\bigcircle{31}}
\drawline[5](-60,2.0)(-182,240)
\drawline[5](-106,2.0)(-229,240)
}
    \caption{ \label{figure1} 
      The B band image of GRB 970111 taken 19 hours after the event. The
      circular field of view has a radius of $8^\prime$. The largest circle
      (radius $3^{\prime}$) represents the $3\sigma$ WFC error circle
      reported by in`t Zand et al.  (1997), later refined by Heise et al.
      (1998) with a radius of $1^{\prime}.8$. The small circle inside the
      WFC error box is the $60^{\prime \prime}$ radius error box
      (confidence level 90\%) given by the NFI for the possible X-ray
      afterglow of GRB 970111. The improved IPN annulus is also shown
      (Galama et al.  1997).  The NFI error circle is only marginally
      consistent with the IPN annulus.  On the other hand, the two
      $3\sigma$ X-ray error boxes sources detected by {\it SAX}, labeled
      {\it SAX} ``a'' and {\it SAX} ``b'' (Butler et al. 1997) are shown.
      North is at the top and east to the left.}
 \end{figure}  

 GRB 970111 was the first GRB observed by {\it BeppoSAX} that was promptly
 followed up at other wavelengths. In fact, the results presented in this
 paper represent the fastest follow-up observation performed for GRB 970111
 in all wavelengths.  It was detected as a three-peak Gamma-Ray Burst on
 January 11, 1997 by the Wide Field Camera of the X-ray {\it BeppoSAX}
 satellite (Costa et al.  1997). The burst was localized at $\alpha =
 15^{h} 28^{m} 24^{s}$, $\delta= +19^{\circ} 40.0^{\prime}$ (equinox
 2000.0; error box radius= $10^{\prime}$).  Soon after that, Butler et al.
 (1997) reported the presence of two faint X-ray sources in the field of
 GRB 970111; 1SAX J1528.8\-+1944 and 1SAX J1528.8\-+1937, labeled as ``a''
 and ``b'' respectively.  Then, Hurley et al. (1997a) reduced the error box
 of GRB 970111 by means of the {\it Ulysses} and BATSE data, and found that
 only the {\it BeppoSAX} source ``a'' lied within it.  Unfortunately, due
 to a misalignment of the Wide Field Cameras of {\it BeppoSAX} (in`t Zand
 et al.  1997), the error box reported by Costa et al. (1997) had to be
 shifted of $\sim 4^{\prime}.2$ from its previous position. The new error
 box provided by Hurley et al.  (1997b), seven times smaller than the
 former, placed both X-ray sources far away ($\sim
 10^{\prime}-15^{\prime}$) from this new error box, thus showing both
 sources to be unrelated to GRB 970111. 1SAX J1528.8+1944 has been detected
 by ROSAT (Voges et al. 1997) and it was found to be related to the
 variable radiosource VLA J1528.7+1945 (Frail et al. 1997). An optical
 spectrum revealed two objects that lied within $1^{\prime\prime}$ from VLA
 J1528.7+1945, showing redshifts $z\sim$0.636 and $z\sim$0.458,
 respectively (Kulkarni et al. 1997). Frontera et al. (1997) provided more
 precise coordinates for this X-ray source, however no object showed
 remarkable variations in the optical, neither in the field of GRB 970111
 nor in the error boxes of the two X-ray sources (Guarnieri et al. 1997b).
 Radio observations at 840 MHz, 1.4 GHz and 1.5 GHz carried out between
 26.4 hours and 120 days did not reveal any steady source in the
 intersection between the WFC $3^{\prime}$ error box with a new improved
 IPN annulus (Galama et al. 1997a, Galama et al. 1997b).  One month after
 the event, the error box was scanned by the BIMA array at 3.5 mm, without
 detecting any fading source at millimeter wavelengths (Smith et al.
 1997).  Very recently the WFC team improved again the error box to an
 irregular circle of $1^{\prime}.8$ radius, still consistent with the
 previous one (Heise et al. 1998).  Furthermore a recent reanalysis of the
 MECS data revealed a previously unknown X-ray source, 1SAX J1528.1\-+1937,
 which is almost entirely contained within the $3\sigma$ WFC error box
 (Feroci et al.  1998).  However this source is only marginally consistent
 with the last IPN annulus reported by Galama et al. (1997b). The above
 mentioned new $\gamma$-ray and X-ray positions of the GRB 970111
 encouraged us to reanalyze the results already presented elsewhere
 (Castro-Tirado et al.  1998b).  Furthermore new observations were carried
 out, which enabled us to observe the GRB 970111 up to 6 months after the
 gamma-ray event.  Section $\S$2 will briefly describe the data reduction
 and calibration techniques, while Section $\S$3 will present and discuss
 the results.  Finally, Section $\S$4 will draw the conclusions.
\begin{table}[t]
\begin{center}
\caption{\label{table1} Log of observations covering the WFC 970111 position.}
  \begin{tabular}{lcccccccr}
    \hline Date & \multicolumn{7}{c}{Exposure time (Ks)} & Telescope \\ 
       of  1997 & U & B & V & R & i & z & free & \\ 
    \hline 
    Jan. 12&- & 1.5 & - &0.6&-  & - &-  &2.2 CAHA   \\ 
    Jan. 17&- & 2.7 & - &0.9&-  & - &-  &1.5 Loiano \\ 
    Feb. 10&- & 1.5 & - &0.6&-  & - &-  &2.2 CAHA   \\ 
    Feb. 11&- & -   & - &0.6&-  & - &-  &2.2 CAHA   \\ 
    Mar.  5&- & -   & - &1.8&-  & - &-  &1.5 Loiano \\ 
    Mar. 10&0.6& 0.6& - &0.6&0.6&0.6&-  &1.54 Danish\\ 
    Mar. 12&- & 1.8 & - &-  &-  & - &-  &1.54 Danish\\ 
    Mar. 14&- & -   & - &-  &-  & - &7.2&1.54 Danish\\ 
    Mar. 15&- & 3.6 & - &-  &-  & - &-  &1.54 Danish\\ 
    Jun. 26&- & -   &4.8&-  &-  & - &-  &1.54 Danish\\ 
    Jun. 27&- & -   & - &3.2&-  & - &-  &1.54 Danish\\ 
    Jul. 1 &- & 2.7 & - &-  &-  & - &-  &1.54 Danish\\ 
    Jul. 3 &- & -   & - &1.8&1.6& - &-  &1.54 Danish\\ 
    Jul. 4 &- & -   & - &-  &2.9& - &-  &1.54 Danish\\ 
    \hline
    \multicolumn{9}{l}{\hspace{1.0cm}0.17$^h$+7.42$^h$+3.37$^h$+4.89$^h$+
      1.42$^h$+0.17$^h$+2$^h$~~~~~~=~19.42$^h$} \\ 
\end{tabular} 
\end{center}
\end{table}

\begin{table}[t]
\begin{center}
\label{table2}
\caption{\label{table2} Log of observations covering the field of source 1SAX J1528.8+1944.}
  \begin{tabular}{lcccccccr}
    \hline 
    Date & \multicolumn{7}{c}{Exposure time (Ks)} & Telescope \\
    of 1997 & U & B & V & R & i & z & free & \\ 
    \hline 
    Jan. 12 & - & 1.5 & - & 0.6 & - & - &-&2.2 CAHA \\ 
    Jan. 14 & - & 1.2 & - & 1.2 & - & - &-&1.5 Loiano \\ 
    Jan. 17 & - & 2.7 & - & 0.9 & - & - &-&1.5 Loiano \\ 
    Jan. 31 & - & 3.0 &1.2 & 0.9 & - & - &-&1.5 Loiano \\ 
    Feb. 10 & - & 1.5 & - &  0.6 & - & - &-&2.2 CAHA \\ 
    Feb. 11 & - & - & - & 0.6 & - & - &-&2.2  CAHA \\ 
    Feb. 14 & - & - &1.8 & - & - & - &-&1.5 Loiano \\ 
    Feb. 17 & - & - &1.62 & 1.8 & - & - &-&1.5 Loiano \\ 
    Feb. 18 & - & 4.5 &2.7 & 1.8 & - & - &-&1.5 Loiano \\ 
    Mar.  5 & - & - & - & 1.8 & - & - &-&1.5 Loiano \\ 
    Mar. 13 & - & 3.6 & - & 1.8 & - & - &-&1.5 Loiano \\ 
\hline

    \multicolumn{9}{c}{\hspace{2.0cm}
      5.0$^h$+2.03$^h$+3.33$^h$~~~~~~~~~~~~~~~~~~~~~~~ ~~~~~~~~=~7.03$^h$}
    \\ 

\end{tabular} 
\end{center}
\end{table}

\begin{table}[t]
\begin{center}

\caption{\label{table3} Log of observations covering the field of source 
 1SAX J1528.8\-+1937.}
  \begin{tabular}{lcccccccr}
    \hline Date & \multicolumn{7}{c}{Exposure time (Ks)} & Telescope \\ 
           of 1997 & U & B & V & R & i & z & free & \\ 
   \hline 
    Jan. 12 & - & 1.5 & - &0.6 & - & - &-&2.2 CAHA \\ 
    Jan. 14 & - & 1.2 & - &1.2 & - & - &-&1.5 Loiano \\ 
    Feb. 10 & - & 1.5 & - &0.6 & - & - &-&2.2 CAHA \\ 
    Feb. 11 & - & - & - &0.6 & - & - &-&2.2 CAHA \\ 
   \hline
    \multicolumn{9}{c}{~~~~~~~~~~~~~~~~~~1.17$^h$~~~~~~+~0.83$^h$~~~
      ~~~~~~~~~~~~~~~~~~~~~~~~~=~2$^h$} \\ 
\end{tabular} 
\end{center}
\end{table}

\begin{table}[t]
\begin{center}

\caption{\label{table4} Filters used in the observations carried out at La Silla.}
  \begin{tabular}{ccccc}
\hline
Filter &  ESO    & Central    & Band-with  & Maximum  \\
name   &  filter & wavelength & FWHM      & transmission \\
       &  number &    (nm)    &  (nm)     &     (\%) \\
\hline
Bessel U & 632 & 355.60 & 53.37  & 66 \\
Bessel B & 450 & 443.57 & 102.29 & 67 \\
Bessel V & 451 & 544.80 & 116.31 & 85 \\
Bessel R & 452 & 648.87 & 164.7  & 85 \\
Gunn   i & 425 & 797.79 & 142.88 & 85 \\
Gunn   z & 462 & 903.35 &        &    \\
\hline
\end{tabular} 
\end{center}
\end{table}

\section{Observations}
Observations were obtained with the 2.2-m telescope at the Calar Alto
Spanish-German observatory (hereafter CAHA), the 1.5-m Telescope of the
Bologna Astronomical Observatory and with the 1.54-m Danish Telescope at La
Silla observatory.  Tables  \ref{table1}, \ref{table2} and \ref{table3}
display the observing logs for the WFC error box (in't Zand et al.  1997),
and the two {\it SAX} X-ray sources labeled ``a'' and ``b''.

\subsection{Observations at CAHA}
We obtained B and R band images on January 12, starting 19 hours the
gamma-ray event. The frames were obtained at the 2.2-m telescope at CAHA 
equipped with the Calar Alto Faint Object Spectrograph (CAFOS). The 
detector used is a 2048 $\times$ 2048 pixel CCD providing a scale of 
0.53 arc sec pixel$^{-1}$. The limiting magnitudes of these
images are B$\sim$23.7, R$\sim$23.0.  

As it is shown in Fig.~\ref{figure1}, the large field of view provided by
CAFOS (16$^{\prime}$ in diameter) enabled us to image all the improved
positions reported in the literature (Butler et al. 1997, Hurley et al.
1997a, 1997b, Galama et al.  1997b, in't Zand et al. 1997, Feroci et al.
1998, Heise et al. 1998) as well as the X-ray sources 1SAX J1528.8+1944 and
1SAX J1528.8\-+1937. Additional comparison B and R frames were taken on
February 10 and 11. We searched for variables objects located inside the
1SAX J1528.8+1944, 1SAX J1528.8\-+1937 and the GRB error box comparing the
images taken in both filters on January 12 with those taken on February 10
and 11. The results are discussed in $\S$ \ref{vari}.

\subsection{Observations at Loiano}

We used the 1.5-m telescope (equipped with the Bologna Faint Object
Spectrograph and Camera, hereafter BFOSC) at the Bo\-logna Astronomical
Observatory, on January 14-15, 17 and 31, on February 14, 17-18 and on
March 5 and 13. The detector used with BFOSC is a 2048 $\times$ 2048 CCD,
with a scale of about 0.5 arc sec pixel$^{-1}$. B and R band filters were
usually used, sometimes V band images were also taken. Frames were also
obtained on February 19, but the seeing was very poor and the frames had a
low S/N ratio; therefore, they were not used in this work.  Long exposures
times were used to reach a limiting magnitude of at least $\sim$ 21 for
each frame.  B,V and R images of the PG 1047+003 sequence (Landolt 1992)
were taken, in order to calibrate the field of GRB 970111.

\begin{table}[b]
\begin{center}
\caption{\label{table5} Magnitudes of the nine secondary standard stars
  shown in Fig.~\ref{figure4}}
  \begin{tabular}{cccccc}
\hline
Star ID. number       & U & B & V & R &i\\
in Fig.~\ref{figure4} &   &   &   &   & \\
\hline
1& 17.52 & 17.49 & 16.86 & 16.47 & 16.11 \\
2& 16.18 & 16.06 & 15.39 & 15.00 & 14.63 \\
3& 16.14 & 15.92 & 15.18 & 14.76 & 14.36 \\ 
4& 17.36 & 17.32 & 16.67 & 16.27 & 15.88 \\
5& 18.04 & 17.32 & 16.37 & 15.81 & 15.29 \\
6& 17.47 & 16.48 & 15.45 & 14.84 & 14.33 \\
7& 16.29 & 16.24 & 15.59 & 15.20 & 14.84 \\
8& 17.18 & 17.29 & 16.82 & 16.48 & 16.15 \\
9& 17.09 & 16.70 & 15.92 & 15.48 & 15.10 \\
\hline
\end{tabular} 
\end{center}
\end{table}

\begin{figure*}[t]
   \centering
\fbox{\resizebox{\hsize}{!}{\includegraphics{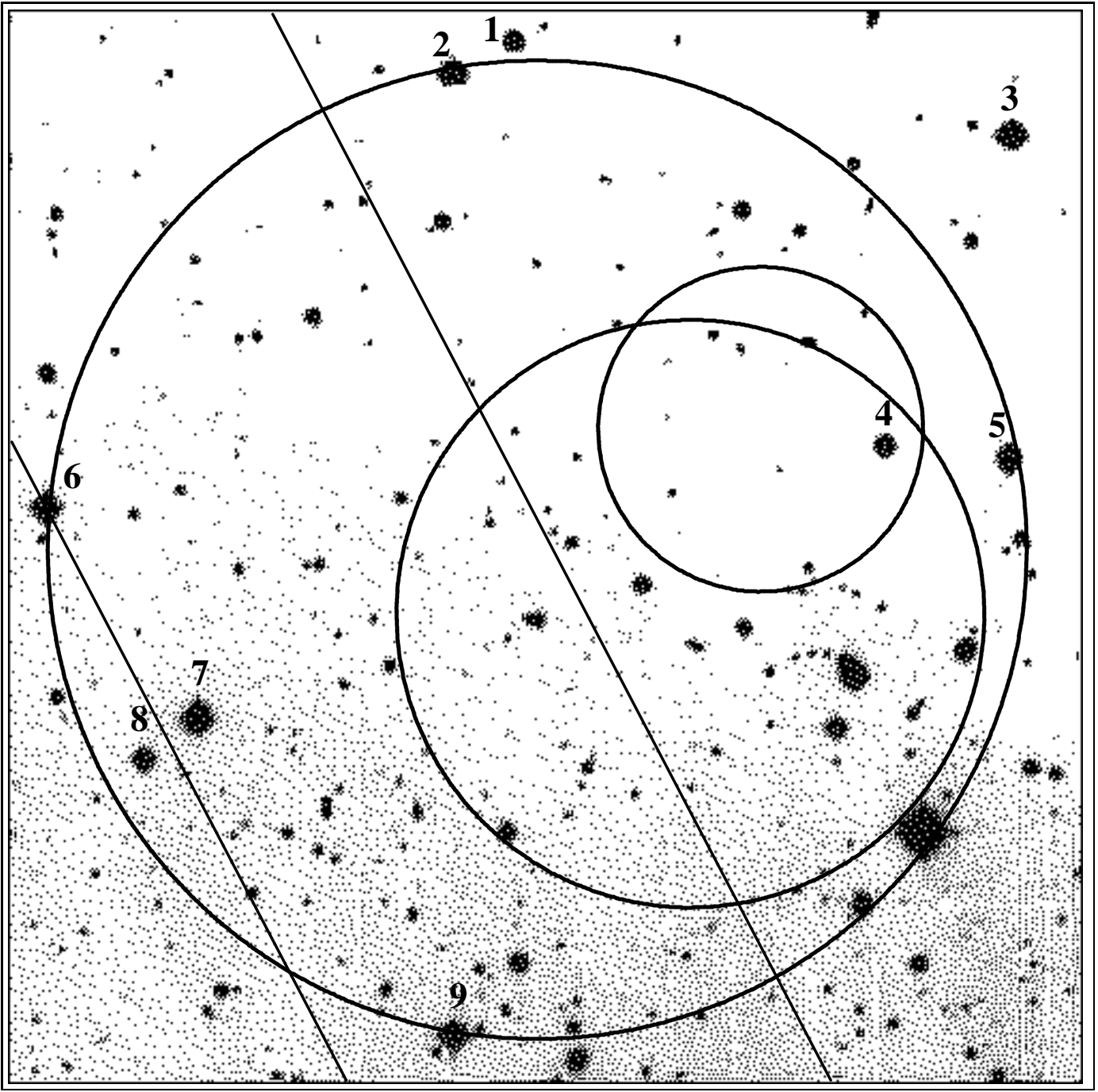}}
\linethickness{0.5mm}
\put(-286,497){\bf \Large 1}
\put(-310,490){\bf \Large 2}
\put(-39,464){\bf \Large 3}
\put(-99,314){\bf \Large 4}
\put(-45,308){\bf \Large 5}
\put(-486,284){\bf \Large 6}
\put(-425,190){\bf \Large 7}
\put(-454,168){\bf \Large 8}
\put(-302,36){\bf \Large 9}
\put(-153.5,312){\bigcircle{155}}
\put(-187,224){\bigcircle{280.3}}
\put(-260,254.5){\bigcircle{467}}
\allinethickness{0.5mm}
\drawline[5](-353.5,1.5)(-513,306)
\drawline[5](-122.5,1.5)(-388.5,510)}
    \caption{Combination of BVR (Bessel) and i (gunn) images taken at La
      Silla for GRB 970111 field. The images includes the two WFC circles
      (in`t Zand et al.  1997, Heise et al. 1998) and the NFI error box
      (Feroci et al. 1998).  The straight lines represent the IPN annulus
      (Galama et al. 1997b).  The total exposure time amounts to 19 hours.
      The field of view is $6^{\prime}.6 \times 6^{\prime}.6$. North is at
      the top and east to the left.}
    \label{figure4}
 \end{figure*}  

\subsection{Observations at La Silla}

The observations were performed during two observing runs on the Danish
1.54-m telescope at ESO La Silla Observatory, the first one in March 1997,
and the second one in June-July 1997. The Instrument used was the Danish
Faint Object Spectrograph and Camera (DFOSC), providing a field of view of
$13^{\prime}.6 \times 13^{\prime}.6$.  The CCD was a backside illuminated
Loral/Lessler chip with 2052 $\times$ 2052 15 $\mu m$ pixels.  Observations
were made using U,B,V,R-Bessel and i,z-gunn filters (see Table
\ref{table4}).

 \begin{figure*}[t]
   \centering
  \includegraphics[height=8.6cm,angle=-89.9]{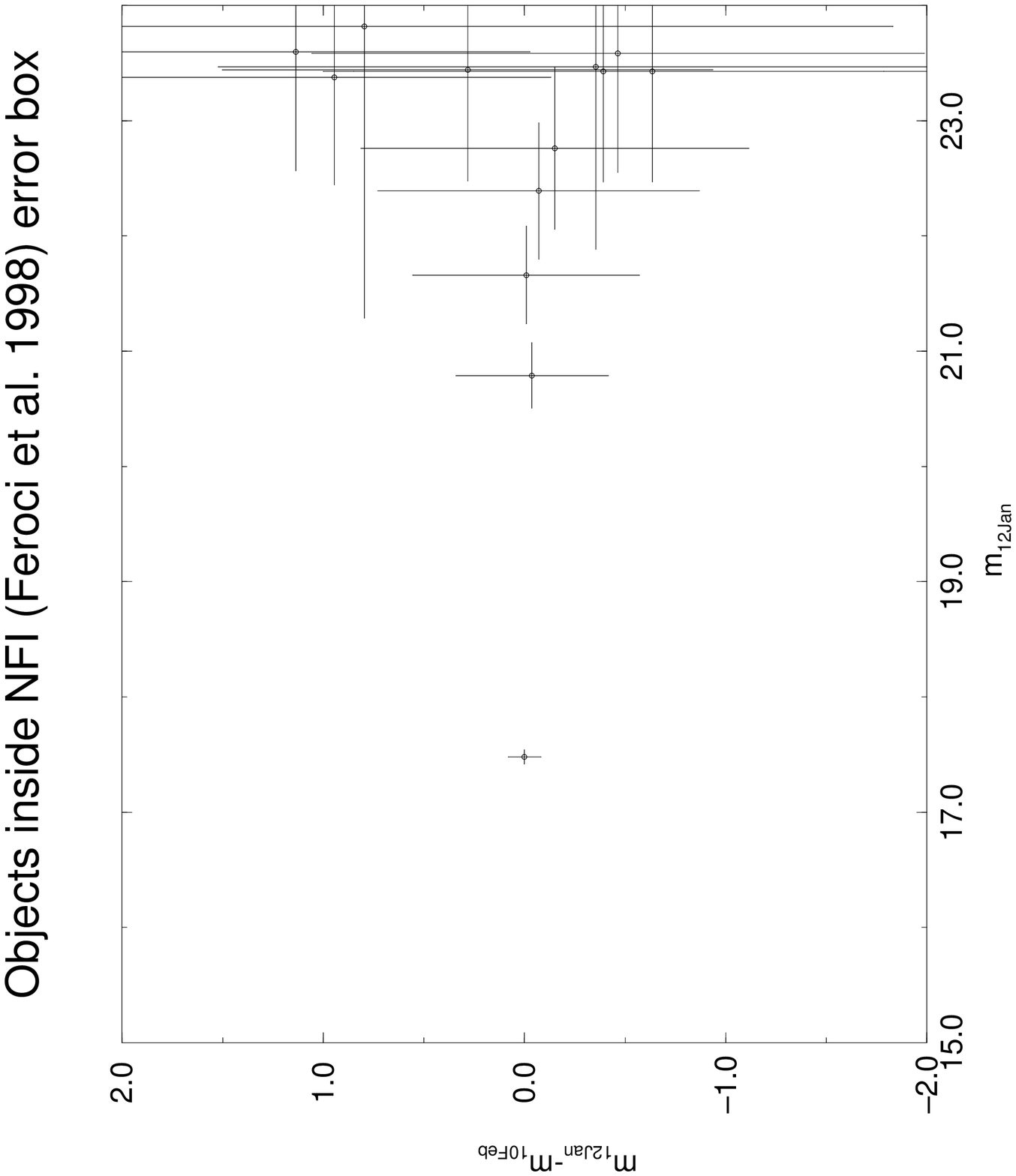}
  \includegraphics[height=8.6cm,angle=-89.9]{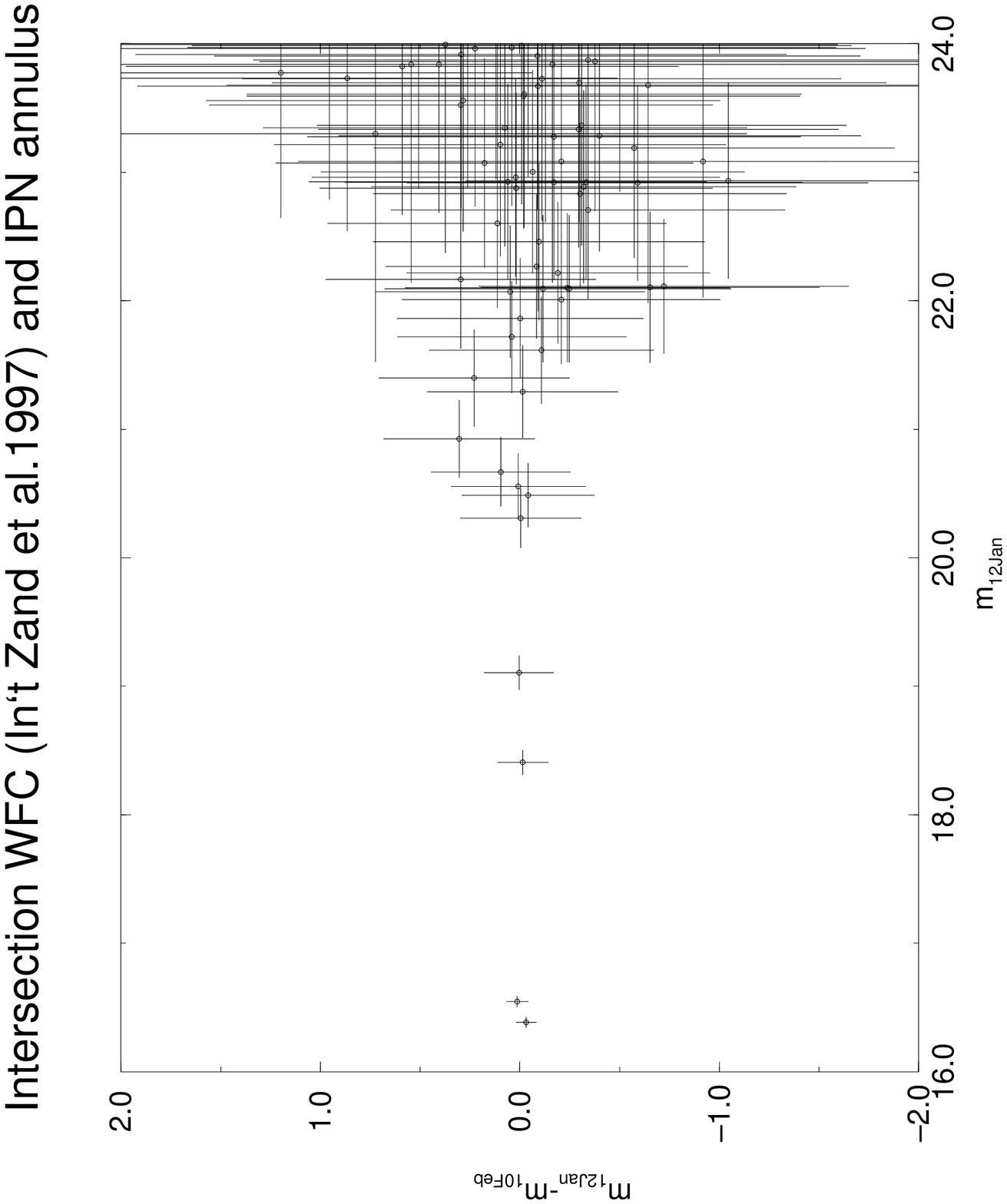}
  \includegraphics[height=8.6cm,angle=-89.9]{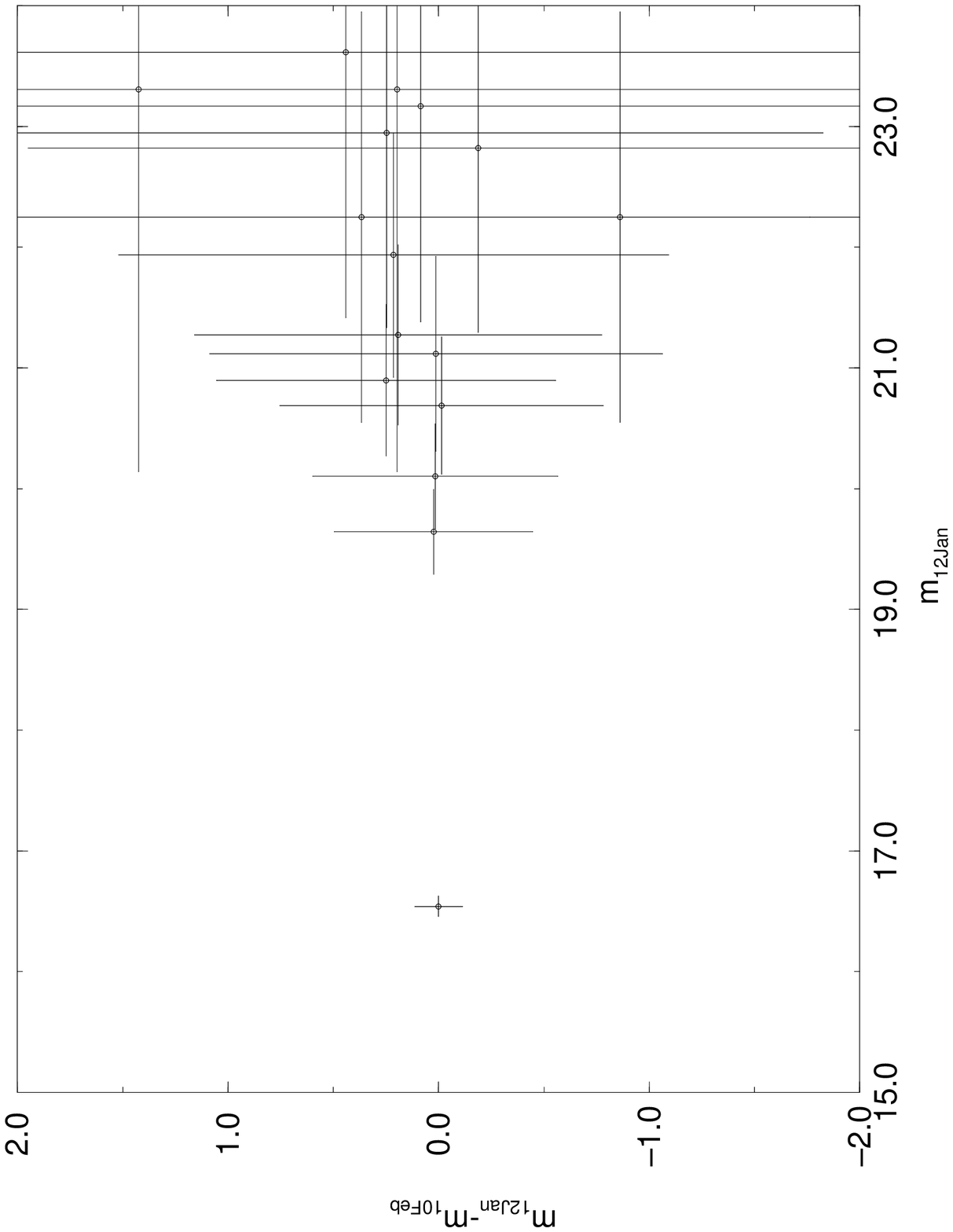}
  \includegraphics[height=8.6cm,angle=-89.9]{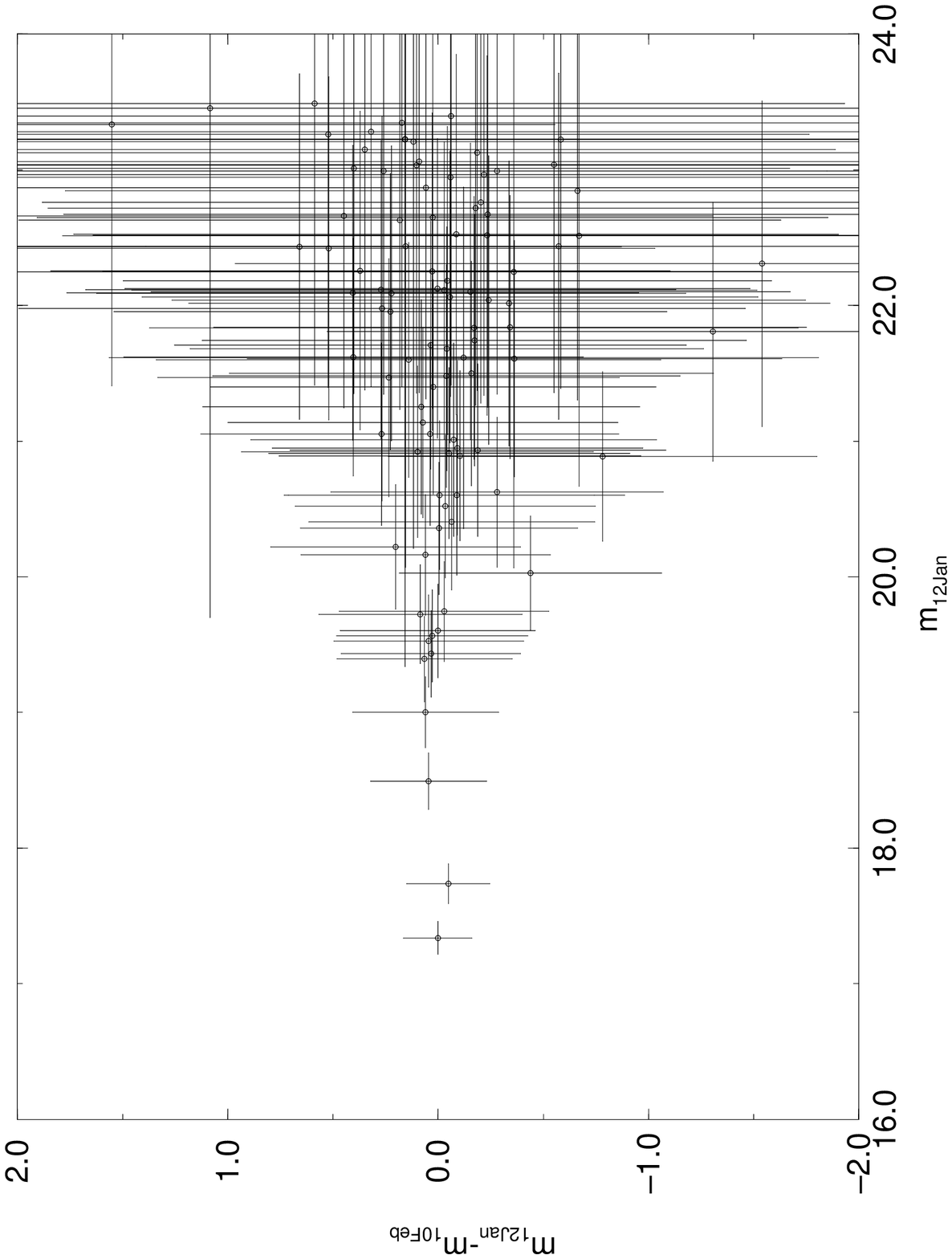}
\caption{    \label{dif} $\Delta$m vs m diagram. The upper two figures show
  the B magnitude difference for the objects in the frames taken with the
  2.2-m Calar Alto telescope 19 hours and 1 month after the gamma-ray
  emission versus the magnitude, for the different error boxes considered.
  The lower two images represent the same diagram for the R filter. As it
  is shown there are no objects varying by more than 1 $\sigma$.}
\end{figure*}

\begin{figure*}[t]
  \fbox{\includegraphics[height=8.7cm,angle=90]{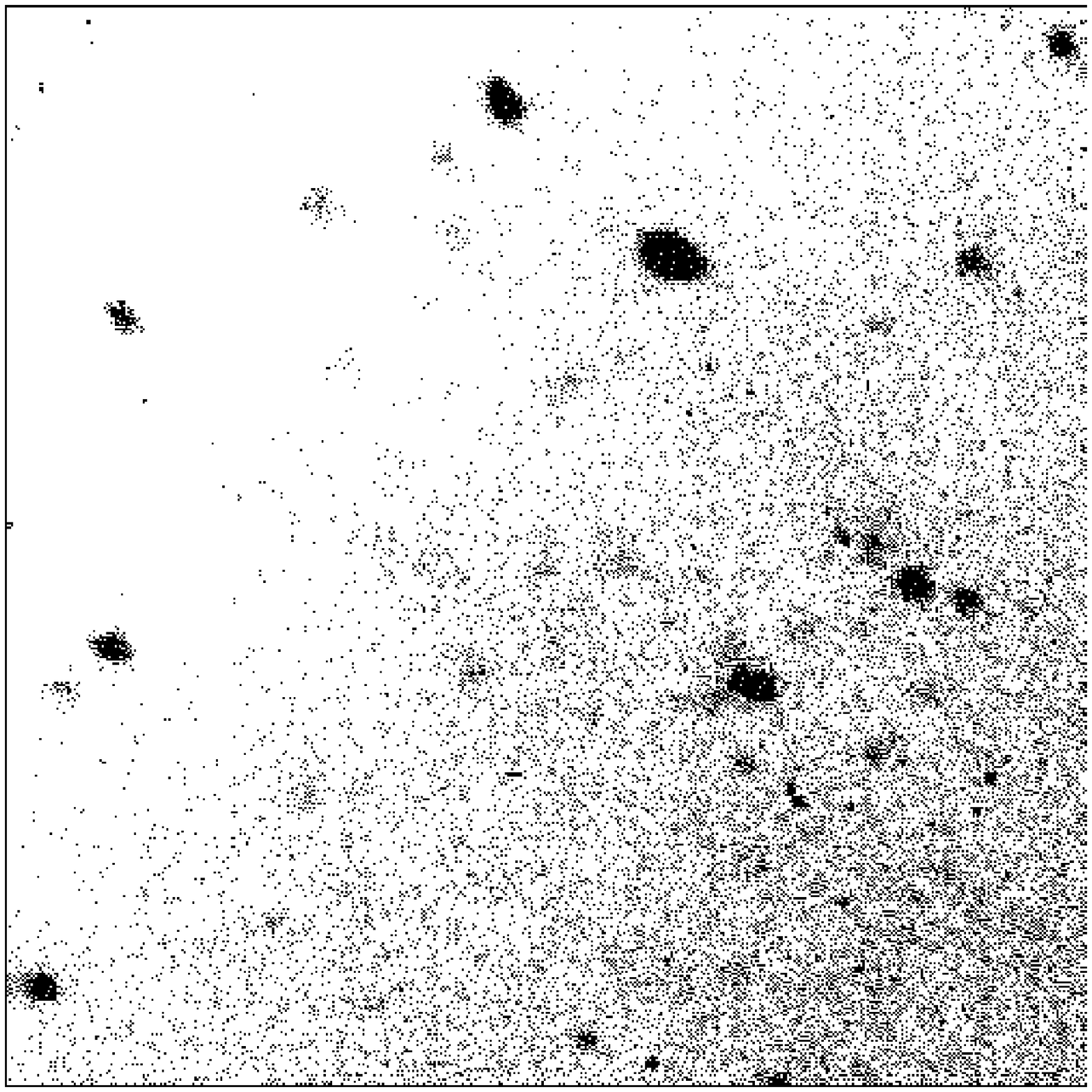}
    \linethickness{0.5mm} \put(-124,124){\bigcircle{245}}
\put(-125,230){\bf 1}
\put(-148,223){\bf 2}
\put(-175,211){\bf 3}
\put(-90,198){\bf 4}
\put(-36,193){\bf 5}
\put(-218,182){\bf 6}
\put(-45,130){\bf 7}
\put(-35,125){\bf 8}
\put(-20,120){\bf 9}
\put(-220,108){\bf 10}
\put(-67,98){\bf 11}
\put(-112,19){\bf 12}
}
\fbox{\includegraphics[height=8.7cm,angle=90]{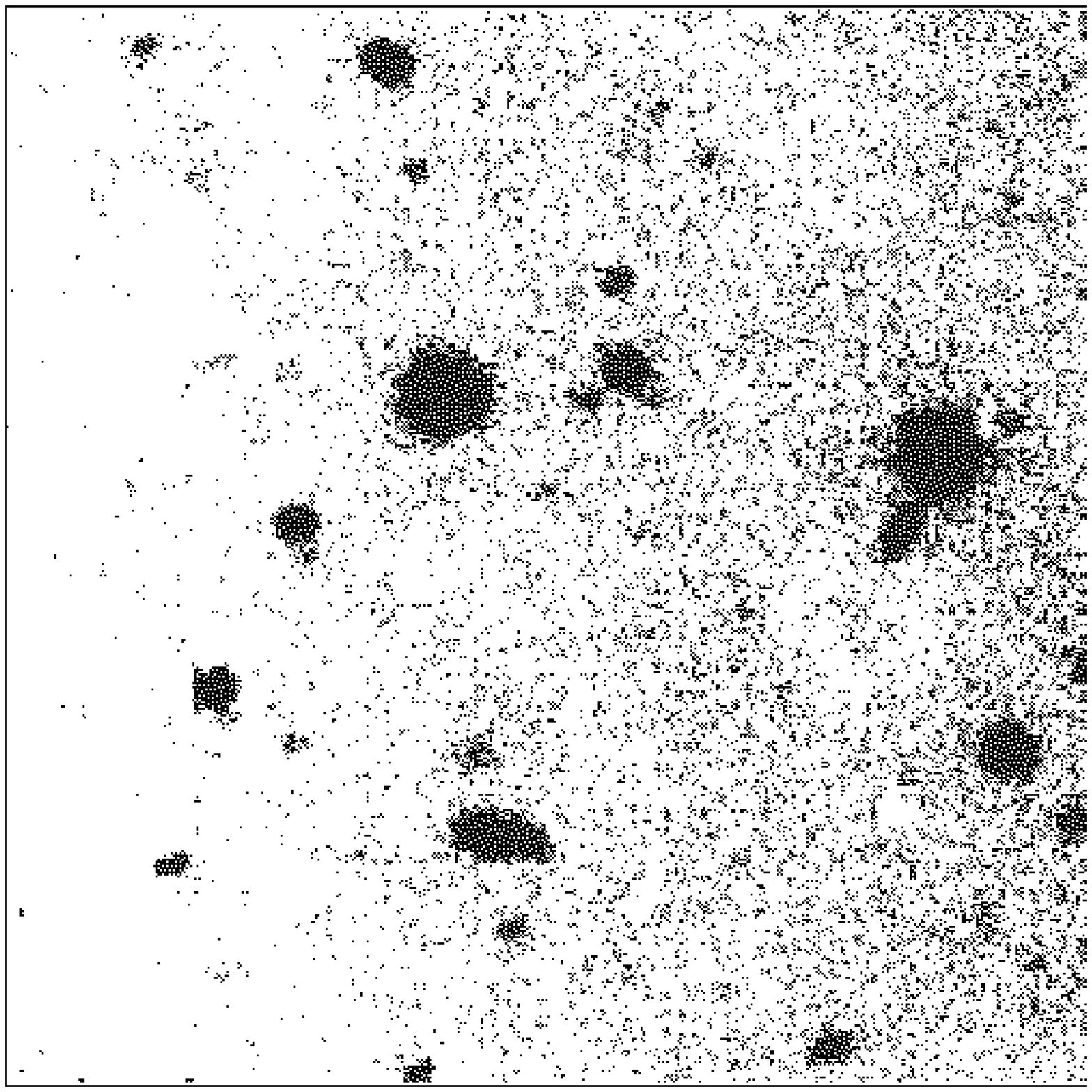}
\linethickness{0.5mm}
\put(-124,124){\bigcircle{246}}
\put(-150,232){\bf 1}
\put(-145,212){\bf 2}
\put(-98,185){\bf 3}
\put(-94,170){\bf 4}
\put(-160,172){\bf 5}
\put(-115,142){\bf 6}
\put(-19,162){\bf 7}
\put(-54,155){\bf 8}
\put(-180,137){\bf 9}
\put(-61,128){\bf 10}
\put(-200,100){\bf 11}
\put(-180,86){\bf 12}
\put(-130,80,){\bf 13}
\put(-31,87){\bf 14}
\put(-120,66){\bf 15}
\put(-205,61){\bf 16}
\put(-122,43){\bf 17}
}

  \caption{   \label{figure2} 
    The error boxes of X-ray sources 1SAX J1528.8+1944 (left) and 1SAX
    J1528.8\-+1937 (right) as seen through the B filter in the images taken
    at Calar Alto on January 12 1997. The field of view is $2^{\prime}
    \times 2^{\prime}$ for both images.  The circles represent the
    $3\sigma$ confidence error boxes with a radii of $1^{\prime}$. The
    magnitudes of the sources inside the X-ray error boxes are displayed in
    Tables \ref{table6} and \ref{table7}. North is at the top and east to
    the left.}
 
\end{figure*}

Table \ref{table5} shows the magnitudes of the nine stars labeled in
Fig.~\ref{figure4}.  Photometric calibrations were performed observing the
Landolt field \#98 (Landolt 1992) under photometric conditions at different
air masses and filters, resulting the following absorption coefficients:
$K_{\rm U}=0.47, K_{\rm B}=0.23, K_{\rm V}=0.13, K_{\rm R} \\ =0.09, K_{\rm
  i}=0.05$. The $1\sigma$ limiting magnitude of the co-added frames are:
U=20, B=24.5, V=24, R=24.5, i=23.

Using these three sets of observations (CAHA, Loiano, and La Silla) we
searched for variable objects inside four different error boxes reported
for GRB 970111; i) the WFC error box with a radius of 3$^{\prime}$ (in`t
Zand et al. 1997), ii) the intersection of i) with the improved IPN annulus
(Galama et al.  1997b), iii) the intersection of the reduced WFC error box
(Heise et al. 1998) with the improved IPN annulus, and iv) the position
given by the NFI error circle for the possible X-ray afterglow (Feroci et
al.  1998).  For clarity Fig.~\ref{dif} only shows the above mentioned
results for ii) and iv).  Colour-colour diagrams were also constructed for
the content of the former regions.  Automatic photometry was carried out
with SEXtractor (Bertin and Arnouts 1996) in the different error boxes
reported for GRB 970111.

\begin{table}[h]
\begin{center}
\label{saxaa}
\caption{\label{table6} Magnitudes for objects inside the X-ray error box of 1SAX J1528.8+1944.}
  \begin{tabular}{ccc}
\hline
&\multicolumn{2}{c}{1SAX J1528.8+1944} \\
\hline 
Object &       B      &     B-R        \\
1    & 21.2   $\pm$0.5&   2.0 $\pm$0.5 \\ 
2    & 23.1   $\pm$1.1&   1.7 $\pm$1.3 \\ 
3    & 22.5   $\pm$0.6&   1.6 $\pm$0.7 \\ 
4    & 20.3   $\pm$0.3&   1.3 $\pm$0.3 \\ 
5    & 22.4   $\pm$0.9&   1.4 $\pm$1.0 \\ 
6    & 22.0   $\pm$0.9&   2.3 $\pm$0.9 \\ 
7    & 22.5   $\pm$0.7&  -0.1 $\pm$0.9 \\ 
8    & 21.9   $\pm$0.6&   1.2 $\pm$0.7 \\ 
9    & 22.4   $\pm$0.8&   2.3 $\pm$0.8 \\
10   & 22.0   $\pm$0.7&   1.2 $\pm$0.8 \\
11   & 21.1   $\pm$0.5&   1.9 $\pm$0.6 \\
12   & 23.4   $\pm$1.2&   1.8 $\pm$1.4 \\
\hline
\end{tabular} 
\end{center}
\end{table}

\begin{table}[h]
\begin{center}
\label{saxbb}
\caption{\label{table7} Magnitudes for objects inside 1SAX J1528.8\-+1937
  X-ray error box.}
  \begin{tabular}{ccc}
\hline
&\multicolumn{2}{c}{1SAX J1528.8\-+1937} \\
\hline 
Object &  B           &  B-R     \\ %
1   &20.0 $\pm$0.2 &2.1 $\pm$0.4 \\ %
2   &23.5 $\pm$1.3 &1.4 $\pm$1.9 \\ 
3   &23.1 $\pm$1.0 &1.3 $\pm$1.1 \\ 
4   &20.6 $\pm$0.3 &0.3 $\pm$0.4 \\ 
5   &19.6 $\pm$0.2 &1.1 $\pm$0.2 \\ 
6   &22.8 $\pm$1.1 &1.7 $\pm$1.2 \\ 
7   &23.2 $\pm$1.2 &2.0 $\pm$1.3 \\ %
8   &17.4 $\pm$0.1 &1.0 $\pm$0.1 \\ 
9   &21.0 $\pm$0.4 &2.1 $\pm$0.4 \\ 
10  &21.8 $\pm$0.6 &1.3 $\pm$0.6 \\ 
11  &21.2 $\pm$0.4 &1.6 $\pm$0.5 \\ 
12  &23.3 $\pm$1.2 &2.3 $\pm$1.2 \\ 
13  &22.5 $\pm$1.1 &0.6 $\pm$1.2 \\ 
14  &19.6 $\pm$0.2 &1.8 $\pm$0.2 \\ %
15  &20.2 $\pm$0.3 &0.8 $\pm$0.3 \\ 
16  &22.5 $\pm$0.9 &0.6 $\pm$1.0 \\ %
17  &22.9 $\pm$1.0 &0.8 $\pm$1.2 \\ 

\hline
\end{tabular} 
\end{center}
\end{table}

\section{Results}
 \subsection{Search for variable objects}
\label{vari}
We compared the images taken at CAHA on January 12 with the ones obtained
on February 10 and 11. Variable sources have neither been found within the
X-ray error boxes of 1SAX J1528.8\-+1944 and 1SAX J1528.8\-+1937, nor in
the whole 16$^{\prime}$ diameter field.

Fig.~\ref{dif} shows the magnitude differences in the B and R filters for
objects inside the error boxes of GRB 970111 when the frames taken in CAHA
19 hours after the burst and $\sim$ 1 month later are compared. If a
suspected variable object is found, it is compared to the images taken at
Loiano and La Silla. As it is shown there is no object varying by more than
$ 1 \sigma$ neither in the R nor in the B filter observations, as reported
elsewhere (Castro-Tirado et al. 1997).  Furthermore, none of the objects
within the different GRB 970111 error box has changed in brightness in the
images taken at Loiano and La Silla.

The fields of both~~X-ray sources~~1SAX J1528.8\-+1944 and 1SAX
J1528.8\-+1937 were observed at Loiano and CAHA using B and R filters.  No
variations in brightness were found for any of the objects within the two
X-ray sources error boxes. Fig.~\ref{figure2} shows the field of both
sources in the B-band.

 \subsection{Search for the optical counterparts of 1SAX J1528.8\-+1944 and 
   1SAX J1528.8\-+1937}

 B and R magnitudes for the objects inside the X-ray error boxes of 1SAX
 J1528.8\-+1944 and 1SAX J1528.8\-+1937 are shown in Tables \ref{table6}
 and \ref{table7}. For objects with magnitudes fainter than 23.5 in B and
 22.5 in R, the errors introduced by the photometry do not allow to get
 reliable values of the B-R colour index. Thus, in Tables \ref{table6} and
 \ref{table7} only objects brighter than the above mentioned magnitudes are
 shown.

 Object \# 11 of Table \ref{table6} is coincident with the radio source VLA
 1528.7+1945 and has been proposed as the optical counterpart of 1SAX
 J1528.8\-+1944 (Frontera et al. 1997, Guarnieri et al. 1997b, Masetti et
 al. 1997a). This object consists of a blend of at least 2 galaxies with
 different redshifts (Kulkarni et al. 1997) and shows a fuzzy and
 complicated structure. The B and R magnitudes reported for object \# 11
 comprises the flux of all its components. This fact explains the
 discrepancy of the B and R magnitudes with those reported by Kulkarni et
 al.  (1997). We also note object \#7 inside error box (Table \ref{table6})
 which shows B-R=-0.1.  However the large error in the colour index makes
 this candidate less likely to be related to 1SAX J1528.8\-+1944.

\begin{figure}[t]
  \includegraphics[height=9cm,angle=-90]{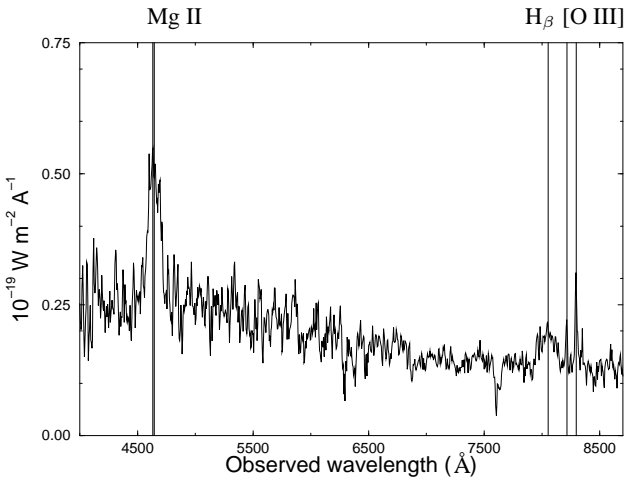}
\put(-87,-195){\AA}
\put(-203,-25){Mg \rm{II}}
\put(-60,-25){H$_{\beta}$}
\put( -46,-25){[O \rm{III}]}

\caption{\label{spectrum}
  The spectrum of object \#4 obtained with 2.2-m (+CAFOS) telescope at
  Calar Alto. The spectrum has been flux calibrated with an instrumental
  response determined in a photometric night a few days later. The redshift
  has been determined to be $z=0.657\pm0.001$ from the narrow emission
  lines of [O III] 4959, 5007. The broad absorption feature at 7610 \AA~is
  the atmospheric A-band.}
\end{figure}

We would like to remark that one of the objects in the 1SAX J1528.8\-+1937
field (object \#4, Table \ref{table7}) is the bluest one with B $<$ 21
found in the $8^{\prime}$ radius CAHA image.  Two optical spectra of this
object were taken with CAFOS at the 2.2-m telescope of CAHA using the B200
grism (3500-6500 \AA) and the R200 grism (6500-10000 \AA). The resolution
is about 12 \AA, given by the $1.25^{\prime \prime}$ slit. Exposure time
was 600 sec in B200 and 1200 sec in R200.

Fig.~\ref{spectrum} shows the spectrum of object \#4, revealing it as a 
typical Seyfert-1 galaxy with broad emission lines of Mg II (2795.5,
2802.7 \AA) and H$_{\beta}$ (4861 \AA), and the narrow emission lines of 
[O III] (4959 \AA, 5007 \AA) observed at wavelengths of 8216 \AA and 
8298 \AA. The broad Mg II line is observed at a central wavelength of
4636 \AA, which is very consistent with the redshift determined from
the narrow lines. The FWHM of the Mg II and H$_{\beta}$ lines imply 
internal velocities of $v\approx 7500$ km s$^{-1}$ and $v \approx 
4600$ km $s^{-1}$, respectively. 

The spectrum was flux calibrated using instrumental responses determined in
a photometric night several days later. Taking into account an interstellar
absorption of $\sim$0.2 mag at $\sim$750 nm derived from the reddening (see
$\S$3.3) and using the flux level taken from the spectrum, we determine the
blue restframe luminosity of this AGN to be between $M_B =$ -21.6 and $M_B
=$ -22.5 depending on cosmology ($q_0 = 0.1$, $H_0 = 75$ km s$^{-1}$
Mpc$^{-1}$ and $H_0 = 50$ km s$^{-1}$ Mpc$^{-1}$ , respectively).

Cumulative surface densities of quasars/\-Seyfert-1 have been determined
observationally by several authors (Hartwick and Schade 1990, Hawkins and
Veron 1995). These authors report the number of quasars and Seyfert-1
galaxies with B$<$21 and $z <$ 2.2 to be about 30 per square degree, rather
independent on the selection criteria employed. Therefore, in our
1$^{\prime}$ radius X-ray error box the {\em a priori} probability for
finding a quasar or Seyfert-1 galaxy with B$<21$ and $z<$ 2.2 is roughly
2.7 \%. Since object \#4 has B=20.6, this probability represents an upper
limit. For this reason, we propose object \#4 as the optical counterpart of
the X-ray source.  Astrometry of object \#4 yields: $\alpha=15^{h} 28^{m}
47.7^{s}$ $ \pm 0.14^{s}, \delta=19^{\circ} 38^{\prime} 53^{\prime\prime}
\pm 2^{\prime\prime}$ (equinox J2000).

\subsection{Colour-colour diagrams}
\label{colour}
\begin{figure*}[t]
  \includegraphics[height=9cm,angle=-89.9]{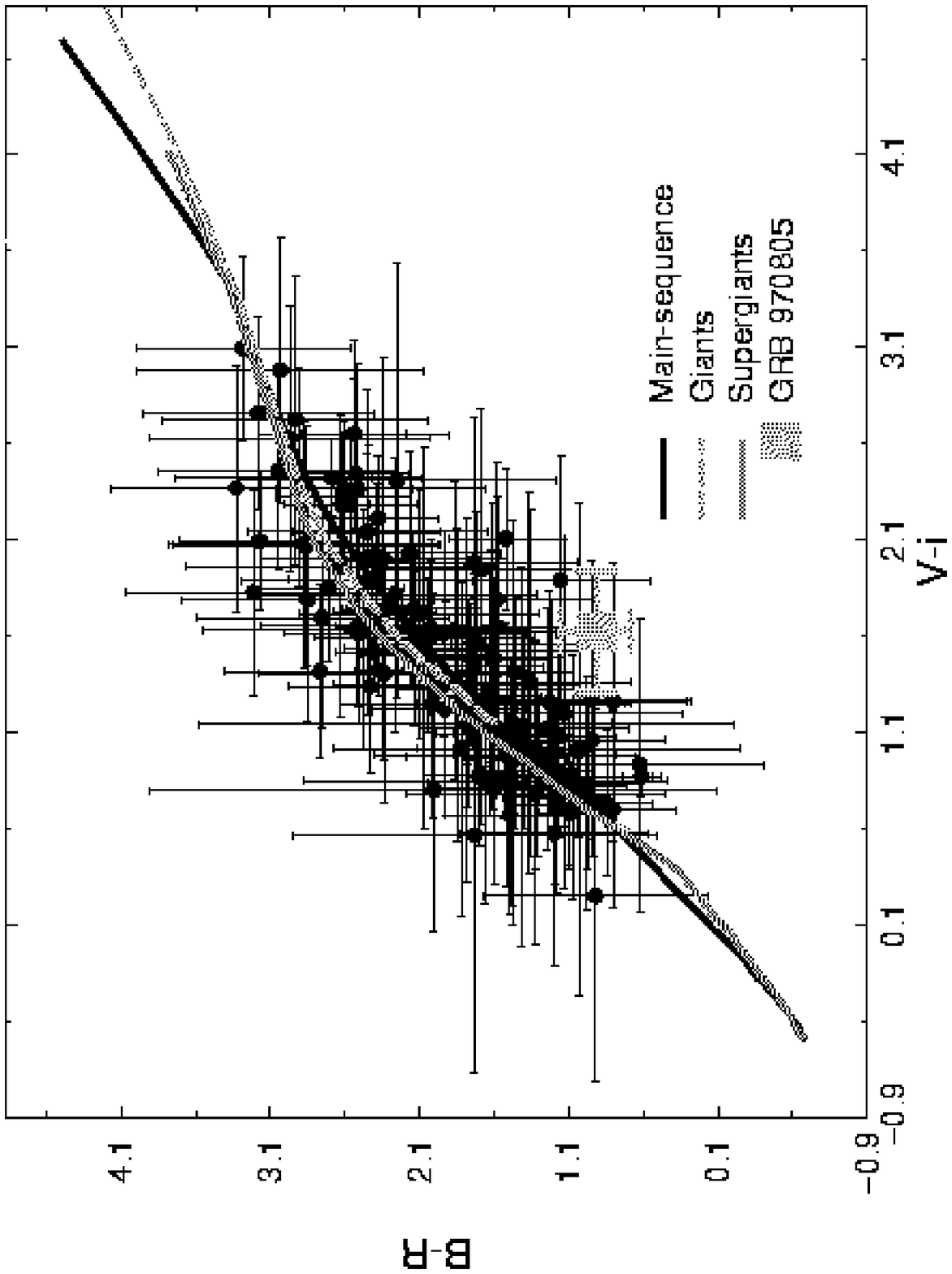}
  \includegraphics[height=9cm,angle=-89.9]{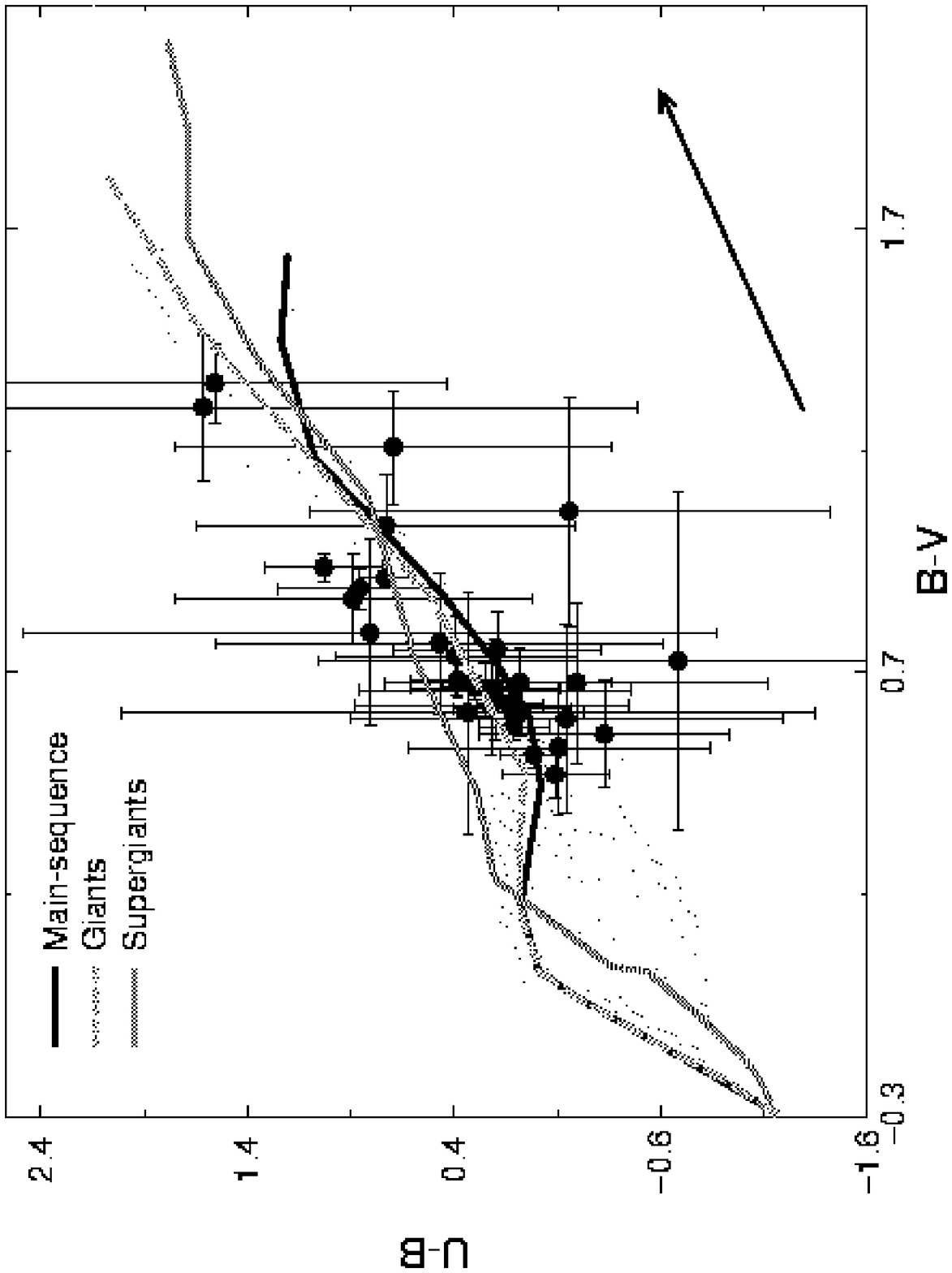}
\caption{  \label{figure6} 
  Colour-colour diagrams for the objects inside the WFC error box. In the
  left-hand panel there is a B-R vs V-i colour-colour diagram for the 157
  objects in the WFC error box for which B,V,R and i measurements were
  possible. As reference, we have plotted ({\it big square}) the colours of
  the GRB 970508 host-galaxy from Zharikov et al.  (1998).  In the
  right-hand panel the U-B vs B-V colour-colour diagram is shown for the 31
  objects.  In both figures the dots represent the colour of real stars
  taken from a photometric catalogue (Lanz 1986).  The different lines show
  the colours expected for main sequence, giant and supergiant stars. The
  arrow indicates the reddening direction.}
\end{figure*}

In order to detect any object with peculiar colours, colour-colour diagrams
were constructed for all objects inside the WFC error box. We consider the
error box reported by in`t Zand et al.  (1997), in order to assure that the
possible error boxes are included.  For 157 objects inside the above
mentioned error box we were able to measure the B, V, R and i magnitudes,
that enabled us to construct a B-R vs V-i diagram (see
Fig.~\ref{figure6}a).  Many objects ($\sim$300) were only detected in the R
and B bands.

In order to distinguish objects with colours not expected in stars and
normal galaxies, we constructed a colour-colour diagram based on the
unreddened colours of 5138 stars of the solar neighbourhood from the
catalogue compiled by Lanz (1986), which we have represented by a dotted
band in the colour-colour diagram. This band is composed by main sequence,
giants and super-giants stars. All the objects with error bars not
overlapping with the band are candidates to high-redshift galaxies,
quasars, active galactic nuclei or whatever object with an optical emission
different from that of normal stars in our galaxy.  We have also
overplotted the colours for GRB 970508 derived from the BVR$_c$I$_c$
photometry performed from Zharikov et al. (1998). The R filter photometric
calibration carried out from Schaefer et al. (1997) and the offsets between
Johnson and Kron-Cousins filters (Frei and Gunn 1994) were taken into
account. As it is shown, the colours of GRB 970508 are clearly away (at
least 2$\sigma$) from the band that represents the main sequence stars.

In spite of the larger errors derived from the observations in the U filter
in comparison to the other filters, we constructed the U-B vs B-V diagram
in order to distinguish objects with ultraviolet excesses, but U,B,V,R and
i magnitudes are available only for 31 objects (see Fig.~\ref{figure6}b).

The uncertainties for the colour indices were calculated from:

$$ \sigma({\rm U-B})=(\sigma(\rm U)^2+\sigma(\rm B)^2)^{1/2}$$
$$ \sigma({\rm B-V})=(\sigma(\rm B)^2+\sigma(\rm V)^2)^{1/2}$$
$$ \sigma({\rm B-R})=(\sigma(\rm B)^2+\sigma(\rm R)^2)^{1/2}$$
$$ \sigma({\rm V-i})=(\sigma(\rm V)^2+\sigma(\rm i)^2)^{1/2}$$

The colour excess E(B-V) of the objects is related to the Hydrogen column
density following Bohlin et al. (1978):

\begin{center}
     E(B-V)$=2.08\times 10^{-22}$ N(H{\sc I}) cm$^{-2}$ \\
\end{center}

The column density along the line of sight of the field $(b^{\rm
  II}=53.4^{\circ}$) is N(H{\sc I})$ \sim 4.8 \times 10 ^{20}$cm$^{-2}$,
which implies a colour excess E(B-V)$\sim$ 0.1.  Once E(B-V) is known the
colour excess for the others filters can be obtained following Cardelli et
al.  (1989), resulting; E(U-B)$\sim$ 0.08, E(B-R)$\sim$0.15,
E(V-i)$\sim$0.13.  Thus the reddening is comparable to the errors on the
photometry.  Furthermore, taking into account that the distance of the
sources are unknown, we will consider the colour-colour diagrams as the
intrinsic colour-colour diagrams for the objects in the field.
Fig.~\ref{figure6} shows both diagrams. As it is shown in the colour-colour
diagrams, there are no sources that deviate over more than a 2$\sigma$
level from the colours expected for the usual components of the galaxies
(main-sequence, giants or supergiants stars).

Among the 31 objects inside the WFC error box for which four colours are
available, there are \ no \ sources \ bluer \ than U-B$<$-0.70, B-R$<$0.60,
B-V $<$0.45, V-R $<$0.32 and R-i $<$0.17.

\subsection{Photometric classification of objects inside the WFC error box,
  and estimation of their redshift}
\label{wolf}

In order to determine the identity of the objects in the WFC error box, we
used a supervised classification method that was originally developed for
the Calar Alto Deep Imaging Survey (CADIS) and is used for object
classification and accurate redshift estimation of galaxies and quasars
there (Wolf et al. 1998a). This method calculates likelihoods for each
individual object to resemble either a star, a galaxy or a quasar, based on
a comparison of its observed colours with libraries of template colours.

These colour libraries have been calculated from spectral libraries by
synthetic photometry based on the UBVRi filterset and the system responses.
As an input, we used the Gunn \& Stryker (1983) catalogue of stellar
spectra, the galaxy template spectra of Kinney et al. (1996) and a quasar
model library derived from the quasar template spectrum of Francis et al.
(1991), which is detailed in Wolf et al. (1998b). Since the galaxy and
quasar libraries are parameterized in redshift, it is also possible to
estimate the redshift for these two classes of objects.

The performance of this supervised multicolour classification strongly
depends on the colour information available and, of course, on the choice of
good libraries. According to Wolf et al. (1998a), this method detects
broad-line AGNs very reliably, and estimates redshifts for galaxies and
quasars, which are mostly accurate to within 5\%, if seven or more filters
are available with photometric accuracies better than 3\%, and the above
mentioned libraries are used. Given only four or five broadband colours,
redshift estimates are bound to be less reliable, but classification will
still distinguish well between stars and galaxies.

As it was explained in $\S$ \ref{colour}, for only 31 objects inside the
WFC error box UBVRi photometry was possible, while for 157 objects only
BVRi magnitudes were measured. In this case, the low number of bands
available and the higher photometric errors of the faint objects left many
objects unclassifiable. For the set of 31 sources, the proportion between
stars, galaxies and unclassifiable objects were: 74\% stars, 18\% galaxies
and 18\% unclassifiable objects. For the set of 157 objects the number of
unclassifiable objects was higher: 27\% stars, 16\% galaxies and 56\% were
not classified. Probably most of the unclassifiable objects in this set are
galaxies because the second set is composed for fainter objects than the
set of 31 objects, and so the fraction between galaxies and stars should be
higher. For twelve objects that were classified as galaxies we were able to 
roughly estimate their redshift, yielding values in the range of $0.2 < z 
< 1.4$. None of the classified objects appears unusual in any respect.

\subsection{Long term red-variable stars}
WX Ser, a Mira-type variable is detected $\sim 5^{\prime}$ outside the GRB
970111 error box. Its period of is 425.1 days and a maximum magnitude of
V=12.0 and a minimum of V$<$16.  However, comparing the B images we taken
on 12-13 March and 15-16 March 1997 to those obtained on July 1997, we
found a variability in the B band of $\Delta$B$ \ge 5$ mag, ranging from
B=20.4$\pm0.1$ to B $<$ 15.4.  On the other hand, the variable object
found by Masetti et al. (1997b) is located outside the field of view of
CAFOS and DFOSC, so we cannot provide further measurements.

\subsection{Similarities between GRB 970111 and GRB 980329}
GRB 980329 is the tenth GRB localized with the WFC on board {\it BeppoSAX}.
The presence of a radio counterpart inside the NFI error box that peaked
$\sim$ 3 days after the gamma-ray event (Taylor et al. 1998), enabled the
detection of a near infrared and an optical (I and R band) counterpart
(Djorgovski et al. 1998c, Klose 1998, Larkin et al. 1998a, 1998b, Palazzi
et al.  1998a, 1998b).

GRB 970111 and GRB 980329 are the most intense GRBs detected by the WFC and
the GRB monitor (GRBM), showing a prominent emission above 40 keV. In fact,
their fluence in the 50-300~keV range is about more than four times larger
than the largest of the other GRBs detected by {\it BeppoSAX}. On the other
hand, they displayed the hardest spectra of the {\it BeppoSAX} GRBs,
showing a hardness ratio (HR) between 0.6 and 0.7 (in't Zand et al. 1998).
Therefore, at first sight one could speculate that both GRBs were
originated under similar physical conditions and nearer than the other GRBs
of the {\it BeppoSAX} sample.

If this was the case, the optical decay curves should be somehow similar.
This fact could give us a clue for explaining the non detectability of GRB
970111 optical transient 19 hours after the gamma-ray event. According to
the power law decay of GRB 980329 ($\alpha$=1.3$\pm$0.2) and the magnitude
of the GRB 980329 optical transient R=23.6$\pm$0.2 20 hours after the
gamma-ray event (Palazzi et al. 1998b), the magnitude 19 hours after the
GRB would be R=23.5, barely detectable in our images of GRB 970111.

Taking into account the fluence of GRB 971214 and assuming that GRBs
resemble standard candles it could imply that both GRBs were originated
from a nearby source in comparison to the source that produced GRB 971214
($z$=3.42). In fact, Palazzi et al. (1998b) suggest that GRB 980329 could
be arised from a strongly obscured starburst galaxy at $z\sim$1.

\section{Conclusion}
The new WFC and NFI positions reported for GRB 970111 (Heise et al. 1998,
Feroci et al. 1998) and the only marginal overlap between the later and the
IPN error box, made us to consider the GRB error box as several non
overlapping regions which had to be analyzed independently. Any possible
fading was $<0.2$ mag for objects with B$<$21, R$<$20.8 and $<0.5$ mag for
those down to B=23 and R=22.6. No fading object was detected, within the
1SAX J1528.8\-+1937 and 1SAX J1528.8+1944 error boxes, being any fading $<$
0.1 mag for objects with B$<$21 and R $<$ 20.8.

The colour-colour diagrams constructed for each non overlapping GRB error
box show neither objects with unusual co\-lours nor objects with ultraviolet
excesses or highly reddened galaxies. The low interstellar extinction in
the direction to GRB 970111 makes the colour-colour diagrams similar to the
unredenned ones. No objects bluer than U-B$<$-0.70, B-R $<$ 0.60, B-V
$<$0.45, V-R $<$0.32, R-i $<$0.17 were found inside the different areas
studied. A photometric classification and redshift estimation of the
objects in the GRB error box revealed no particularly conspicuous object.

On the basis of the the B-R index we have found a possible candidate with
B=20.6 for the X-ray source detected by SAX, called 1SAX J1528.8\-+1937
which is the bluest object found in the $8^{\prime}$ radius image.
Spectroscopic observations revealed that it is a Seyfert-1 galaxy at
redshift $z=0.657$. According to previous surveys, the a {\it priori}
probability that a quasar or Seyfert-1 galaxy with $ z <$ 2.2 and B$<$21 is
located by chance inside the 1$^{\prime}$ X-ray error box, is at most
$\sim$ 3\%.  Therefore, we propose the Seyfert-1 galaxy as the source of
1SAX J1528.8\-+1937.

On other hand, we report the large amplitude found in the variable star WX
Ser ($\Delta$B $\ge$ 5 mag) located $\sim 5^{\prime}$ outside the GRB
970111 error box.

GRB 970111 shows similar characteristics to GRB 980329, being the most
intense and showing the hardest spectra of the GRBs detected by {\it
  BeppoSAX}.  If they also shared a similar optical decay, the magnitude 19
hours after the GRB would be R=23.5 (barely detectable in our images of GRB
970111).  This fact could explain the lack of detection of the GRB 970111
optical transient.

\section*{Acknowledgments}
We are grateful to R. Castillo for the help provided at La Silla and to M.
de Santos-Lle\'o for fruitful discussions.  This work has been partially
supported by Spanish CICYT grant ESP95-0389-C02-02 and by the University of
Bologna (Funds for selected research topics). Jochen Heidt and Thomas Seitz
acknowledge support by the Deutsche Forschungsgemeinschaft through SFB 328.
\vfill
\eject

\end{document}